# Universal scaling relation in high-temperature superconductors

C. C. Homes[1], S. V. Dordevic[1], M. Strongin[1], D. A. Bonn[2], Ruixing Liang[2], W. N. Hardy[2], Seiki Koymia[3], Yoichi Ando[3], G. Yu[4], X. Zhao[5], M. Greven[5,6], D. N. Basov[7], and T. Timusk[8]

[1]*Department of Physics, Brookhaven National Laboratory, Upton, New York 11973*

[2]*Department of Physics and Astronomy, University of British Columbia, Vancouver, B.C. V6T 2A6, Canada*

[3]*Central Research Institute of Electric Power Industry, Komae, Tokyo 201-8511, Japan*

[4]*Department of Physics, Stanford University, Stanford, CA 94305*

[5]*Stanford Synchrotron Radiation Laboratory, Stanford, CA 94309*

[6]*Department of Applied Physics, Stanford University, Stanford, California 94305*

[7]*Department of Physics, University of California at San Diego, La Jolla, California 92093*

[8]*Department of Physics and Astronomy, McMaster University, Hamilton, Ontario, Canada L8S 4M1*

**Scaling laws express a systematic and universal simplicity among complex systems in nature. For example, such laws are of enormous significance in biology[1]. Scaling relations are also important in the physical sciences. The seminal 1986 discovery[2] of high transition-temperature (high-$T_c$) superconductivity in cuprate materials has sparked an intensive investigation of these and related complex oxides, yet the mechanism for superconductivity is still not agreed upon. In addition, no universal scaling law involving such fundamental properties as $T_c$ and the superfluid density $\rho_s$, a quantity indicative of the number of charge carriers in the superconducting state, has been discovered. Here we demonstrate that the scaling relation $\rho_s \propto \sigma_{dc}T_c$, where the conductivity $\sigma_{dc}$ characterizes the unidirectional, constant flow of electric charge carriers just above $T_c$, universally holds for a wide variety of materials and doping levels. This surprising unifying observation is likely to have important consequences for theories of high-$T_c$ superconductivity.**

Since the discovery of superconductivity at elevated temperatures in the cuprate materials there has been a considerable effort to find universal trends and correlations among physical quantities. One of the earliest patterns that emerged was the linear scaling of the superfluid density $\rho_s = c^2/\lambda^2$ ($c$ is the speed of light and $\lambda$ is the penetration depth) with the superconducting critical temperature $T_c$, which marks the onset of phase coherence. This is referred to as the Uemura relation[3], and it works reasonably well for the underdoped materials. However, it does not describe optimally doped (i.e., $T_c$ is a maximum) or overdoped materials[4], or the electron-doped systems[5,6]. It has been observed that in the high-$T_c$ superconductors, a large fraction of the normal-state carriers participate in the superconductivity[7], thus $\rho_s \propto n_s/m^*$ should be sensitive to the $n/m^*$ of the normal state, regardless of $T_c$ ($n$ is the carrier concentration, and $m^*$ is the effective mass). In this sense, it is not surprising that the Uemura relation can not describe $\rho_s$ based strictly on $T_c$ alone. Similarly, an attempt to scale $\lambda$ with $\sigma_{dc}$ was only partially successful[8]. We demonstrate here that the simple relation $\rho_s \propto \sigma_{dc} T_c$, with $\sigma_{dc}$ measured at $T \cong T_c$, universally holds for all high-$T_c$ materials, regardless of doping level, crystal structure and type of disorder[9], nature of dopant (electrons versus holes) and direction (parallel or perpendicular to the copper oxygen planes). Moreover, we show that this relation approximately holds for some conventional superconductors as well.

We first demonstrate scaling for the *ab*-plane (parallel to the copper-oxygen planes) properties[10-16] of single and double-layer cuprates (Table 1), as well as for the conventional metals Nb and Pb (elemental superconductors with relatively high $T_c$'s). The values for $\rho_s$ and $\sigma_{dc}$ are obtained simultaneously from studies of the reflectance of these materials. The reflectance is a complex quantity consisting of amplitude and a phase; in an experiment only the amplitude is usually measured. However, if the reflectance is measured over a wide frequency range, the Kramers-Kronig relation may be used to obtain the phase. Once the complex reflectance is known, then other complex optical functions may be calculated (e.g., the complex dielectric function). The results for the scaling relation are shown on a log-log plot in Fig. 1. The dashed line is a linear fit to the data, while the dotted lines form what is effectively an upper and lower bound for the data; this is described by $\rho_s = (120 \pm 25)\ \sigma_{dc}\ T_c$ (where $\rho_s$ is in $cm^{-2}$, $\sigma_{dc}$ is in $\Omega^{-1}cm^{-1}$, and $T_c$ is in K). The powerful result contained in this plot is that within error all of these points fall onto a single line with a slope of unity. This is significant, as the optimally and overdoped materials, which fell well off of the Uemura plot, now scale with the underdoped materials onto a single line.

We also explored scaling relations along the poorly-conducting *c*-axis, where the charge transport is thought to be incoherent[17]. Previous work focused on scaling between $\rho_s$ and $\sigma_{dc}$ only[18,19]. While this approach yields reasonable results for the underdoped materials, in a fashion that is reminiscent of the Uemura plot, significant deviations from linear behavior are encountered for optimally and overdoped materials; this was thought to signal the onset of more conventional three-dimensional behavior. Figure 2 demonstrates that the *c*-axis data[19-22] for all of the single and double-layer materials (Table 2) are again well described by a line with slope of unity. What is perhaps most remarkable is that the *ab*-plane and *c*-axis results may all be described by the same universal line shown in Fig. 2, even though the two results correspond to different ranges of $\rho_s$.

The scaling relation for the *a-b* planes can be interpreted in a number of different ways. One of the most direct is the assumption that all of the spectral weight associated with the free-carriers collapses into the superconducting condensate ($n_s \equiv n$) below $T_c$. Allowing that the low-frequency conductivity at $T \cong T_c$ can be described by a simple Drude response, $\sigma_1(\omega) = \sigma_{dc}/(1+\omega^2\tau^2)$, which has the shape of a Lorentzian centered at zero frequency with a width at half-maximum of the scattering rate $1/\tau$, the area under this curve may be approximated simply as $\sigma_{dc}/\tau$. Transport measurements for the cuprates[23] suggest that $1/\tau$ near the transition scales with $T_c$, so the strength of the condensate is just $\rho_s \propto \sigma_{dc}\, T_c$, in agreement with the observed scaling relation. This result requires that these materials approach the clean limit ($1/\tau << 2\Delta$, assuming an isotropic superconducting energy gap $2\Delta$).

However, this approach can not be applied to the properties along the *c* axis, where it is generally conceded that transport in this direction is incoherent, and therefore hopping rather than scattering governs the physics[17]. The two-dimensional nature of the cuprates, which often includes a semiconducting or activated response of the resistivity along the *c* axis, has resulted in the description of the superconductivity in this direction in terms of a Josephson-coupling picture[18,19,24-27]. The *c*-axis penetration depth is then determined by the Josephson current density $J_c$ and is $\lambda^2 = \hbar c^2/8\pi d e J_c$, where $d$ is the separation between the planes[25]. There is convincing evidence that the energy gap in the cuprates is *d*-wave in nature, containing nodes at the Fermi surface[28-30]. Calculating $J_c$ for a *d*-wave superconductor is difficult and beyond the scope of this Letter.

This new scaling relation allows the prediction of the penetration depth from measurements of $T_c$ and the normal state conductivity. This scaling relation, which applies within the copper-oxygen planes as well as perpendicular to them, is followed over five orders of magnitude from the insulating behavior along the *c* axis in the underdoped systems to the metallic behavior in the *a-b* planes of the overdoped cuprates. The robust nature of this relation should serve as a guide to establish a new level of understanding of the superconductivity in the cuprates.

1. (Brown, J. H. & West, G. B., eds). *Scaling in biology*. Oxford University Press, Oxford, (1999).

2. Bednorz, J. G. & Mueller, K. A. Possible high $T_c$ superconductivity in the Ba-La-Cu-O system. *Z. Phys. B* **64**, 189–193 (1986).

3. Uemura, Y. J. *et al.* Universal Correlations between $T_c$ and $n_s/m^*$ (Carrier Density over Effective Mass) in High-$T_c$ Cuprate Superconductors. *Phys. Rev. Lett.* **62**, 2317–2320 (1989).

4. Niedermayer, C. *et al.* Muon spin rotation study of the correlation between $T_c$ and $n_s/m^*$ in overdoped $Tl_2Ba_2CuO_{6+\delta}$. *Phys. Rev. Lett.* **71**, 1764–1767 (1993).

5. Homes, C. C., Clayman, B. P., Peng, J. L. & Greene, R. L. Optical properties of $Nd_{1.85}Ce_{0.15}CuO_4$. *Phys. Rev. B* **56**, 5525–5534 (1997).

6. Singley, E. J., Basov, D. N., Kurahashi, K., Uefuji, T. & Yamada, K. Electron dynamics in $Nd_{1.85}Ce_{0.15}CuO_{4+\delta}$: Evidence for the pseudogap state and unconventional *c*-axis response. *Phys. Rev. B* **64**, 224503 (2001).

7. Tanner, D. B. *et al.* Superfluid and normal-fluid densities in high-$T_c$ superconductors. *Physica B* **244**, 1–8 (1998).

8. Pimenov, A. *et al.*, Universal relationship between the penetration depth and the normal-state conductivity in YBCO. *Euorphys. Lett.* **48**, 73–78 (1999).

9. Eisaki, H. *et al.,* Effect of chemical inhomogeneity in bismuth-based copper oxide superconductors. *Phys. Rev. B* **69**, 064512 (2004).

10. Basov, D. N. *et al.* In-Plane Anisotropy of the Penetration Depth in $YBa_2Cu_3O_{7-x}$ and $YBa_2Cu_4O_8$ Superconductors. *Phys. Rev. Lett.* **74**, 598–601 (1995).

11. Homes, C. C. *et al.* Effect of Ni impurities on the optical properties of $YBa_2Cu_3O_{6+x}$. *Phys. Rev. B* **60**, 9782–9792 (1999).

12. Liu, H. L. *et al*. Doping-induced change of optical properties in underdoped cuprate superconductors. *J. Phys: Condens. Matter* **11**, 239–264 (1999).

13. Puchkov, A. V., Timusk, T., Doyle, S. & Herman, A. M. *ab*-plane optical properties of $Tl_2Ba_2CuO_{6+\delta}$. *Phys. Rev. B* **51**, 3312–3315 (1995).

14. Startseva, T. *et al.* Temperature evolution of the pseudogap state in the infrared response of underdoped $La_{2-x}Sr_xCuO_4$. *Phys. Rev. B* **59**, 7184–7190 (1999).

15. Pronin, A. V. *et al.* Direct observation of the superconducting energy gap developing in the conductivity spectra of niobium. *Phys. Rev. B* **57**, 14416–14421 (1998).

16. Klein, O., Nicol, E. J., Holczer, K. & Grüner, G. Conductivity coherence factors in the conventional superconductors Nb and Pb. *Phys. Rev. B* **50**, 6307–6316 (1994).

17. Ando, Y. *et al.*, Metallic in-plane and divergent out-of-plane resistivity of a High-$T_c$ cuprate in the zero temperature limit. *Phys. Rev. Lett.* **77**, 2065-2068 (1996).

18. Dordevic, S. V. *et al.* Global trends in the interplane penetration depth of layered superconductors. *Phys. Rev. B* **65**, 134511 (2002).

19. Basov, D. N., Timusk, T., Dabrowski, B. & Jorgensen, J. D. *c*-axis response of $YBa_2Cu_4O_8$: A pseudogap and possibility of Josephson coupling of $CuO_2$ planes. *Phys. Rev. B* **50**, 3511–3514 (1994).

20. Homes, C. C., Timusk, T., Bonn, D. A., Liang, R. & Hardy, W. N. Optical properties along the *c* axis $YBa_2Cu_3O_{6+x}$, for $x = 0.50 \rightarrow 0.95$: Evolution of the pseudogap. *Physica C* **254**, 265–280 (1995).

21. Schützmann, J., Tajima, S., Miyamoto, S. & Tanaka, S. *c*-Axis Optical Response of Fully Oxygenated $YBa_2Cu_3O_{7-\delta}$: Observation of Dirty-Limit-Like Superconductivity and Residual Unpaired Carriers. *Phys. Rev. Lett.* **73**, 174–177 (1994).

22. Basov, D. N. *et al.* Sum rules and interlayer conductivity of High-$T_c$ cuprates. *Science* **283**, 49–52 (1999).

23. Orensetein, J. *et al.* Frequency- and temperature-dependent conductivity in $YBa_2Cu_3O_{6+x}$ crystals. *Phys. Rev. B* **42**, 6342–6362 (1990).

24. Shibauchi, T. *et al.* Anisotropic penetration depth in $La_{2-x}Sr_xCuO_4$. *Phys. Rev. Lett.* **72**, 2263–2266 (1994).

25. Lawrence, W. E. & Doniach, S. in *Proceedings of the 12th International Conference on Low Temperature Physics*, (Kando, E., ed), 361 (Academic Press, Kyoto, 1971).

26. Bulaevskii, L. N. Magnetic properties of lamellar superconductors with weak interaction between the layers. *Sov. Phys. JETP* **37**, 1133–1139 (1973).

27. Ambegaokar, V. & Barato., A. Tunneling Between Superconductors. *Phys. Rev. Lett.* **10**, 486–489 (1963).

28. Hardy, W. N., Bonn, D. A., Morgan, D. C., Liang, R. & Zhang, K. Precision measurements of the temperature dependence of lambda in $YBa_2Cu_3O_{6.95}$: Strong evidence for nodes in the gap function. *Phys. Rev. Lett.* **70**, 3999–4002 (1993).

29. Ding, H. *et al.* Angle-resolved photoemission spectroscopy study of the superconducting gap anisotropy in $Bi_2Sr_2CaCu_2O_{8+x}$. *Phys. Rev. B* **54**, R9678–R9681 (1996).

30. Damascelli, A., Hussain, Z. & Shen, Z.-X. Angle-resolved photoemission studies of the cuprate superconductors. *Rev. Mod. Phys.* **75**, 473–541 (2003).

**Acknowledgements**

The authors would like to thank A. Chubukov, P. D. Johnson, S. A. Kivelson, P. A. Lee, D. B. Tanner, J. J. Tu, Y. Uemura and T. Valla for useful discussions. Work in Canada was supported by the Natural Sciences and Engineering Research Council of Canada, the Canadian Institute for Advanced Research. The $HgBa_2CuO_{4+\delta}$ crystal growth work at Stanford University was supported by the Department of Energy's Office of Basic Energy Sciences, Division of Materials Sciences and Engineering. Work the University of California at San Diego was supported by the National Science Foundation and DOE. Work at Brookhaven was supported by the DOE.

**Correspondence and requests for materials should be addressed to C.C.H. (e-mail: homes@bnl.gov).**

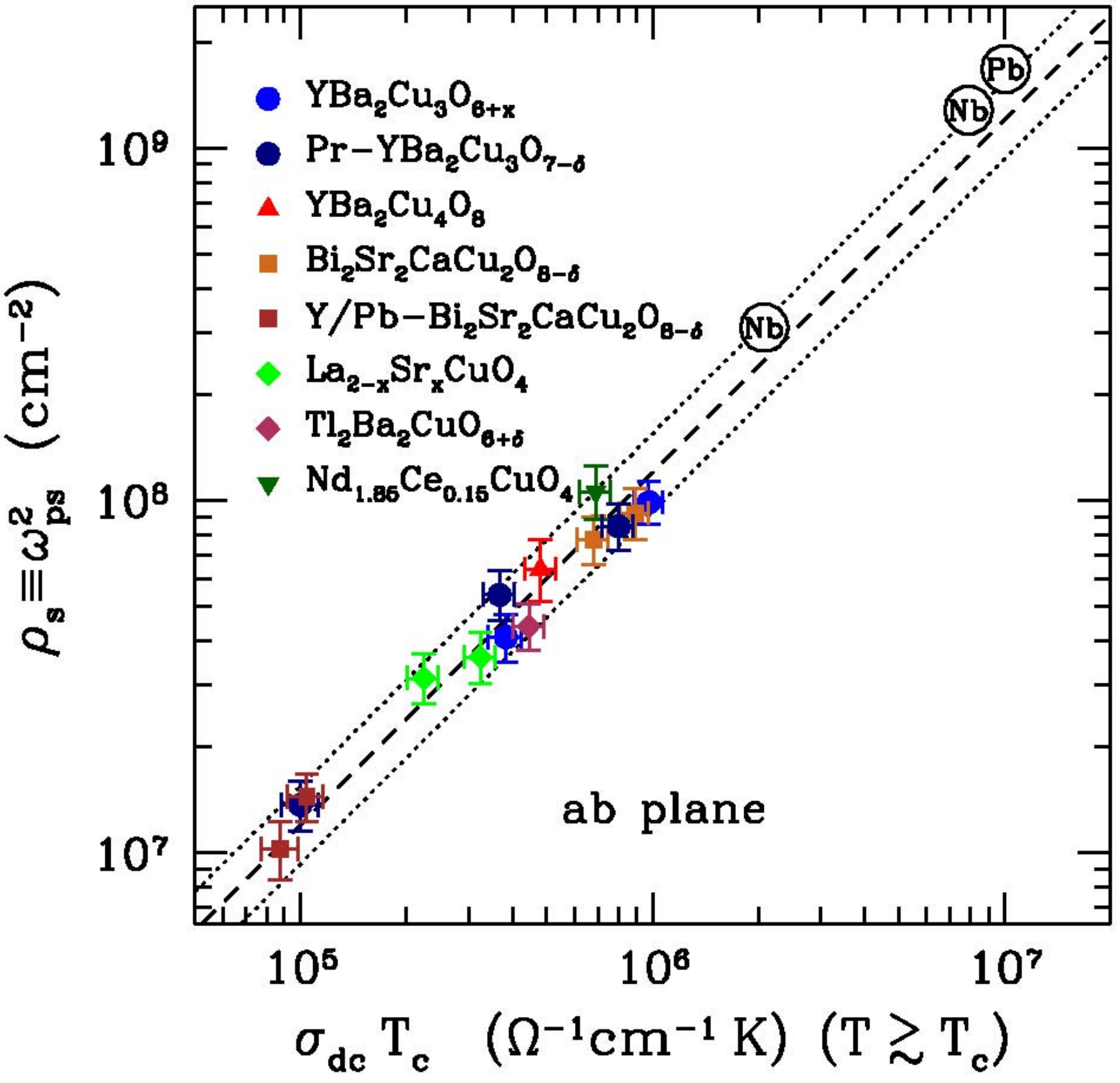

Figure 1: The log-log plot of the superfluid density $\rho_s \equiv \omega_{ps}{}^2$ vs $\sigma_{dc}$ $T_c$ parallel to the copper-oxygen (*a*-*b*) planes for a variety of cuprates as well as some simple metals (Table 1). The dc conductivity used in this scaling relation has been extrapolated from the optical conductivity $\sigma_{dc} = \sigma_1(\omega \to 0)$ at $T \cong T_c$. For $T << T_c$, the response of the dielectric function to the formation of a condensate is expressed purely by the real part $\varepsilon_1(\omega) = \varepsilon_\infty - \omega_{ps}{}^2/\omega^2$, which allows the strength of the condensate to be calculated from $\rho_s = -\omega^2 \varepsilon_1(\omega)$ in the $\omega \to 0$ limit. The dashed and dotted lines are described by $\rho_s = (120 \pm 25)\ \sigma_{dc}\ T_c$. Within error, all the data for the cuprates are described by the dashed line. The data for the conventional superconductors Nb and Pb, indicated by the atomic symbols within the circles, lie slightly above the dashed line.

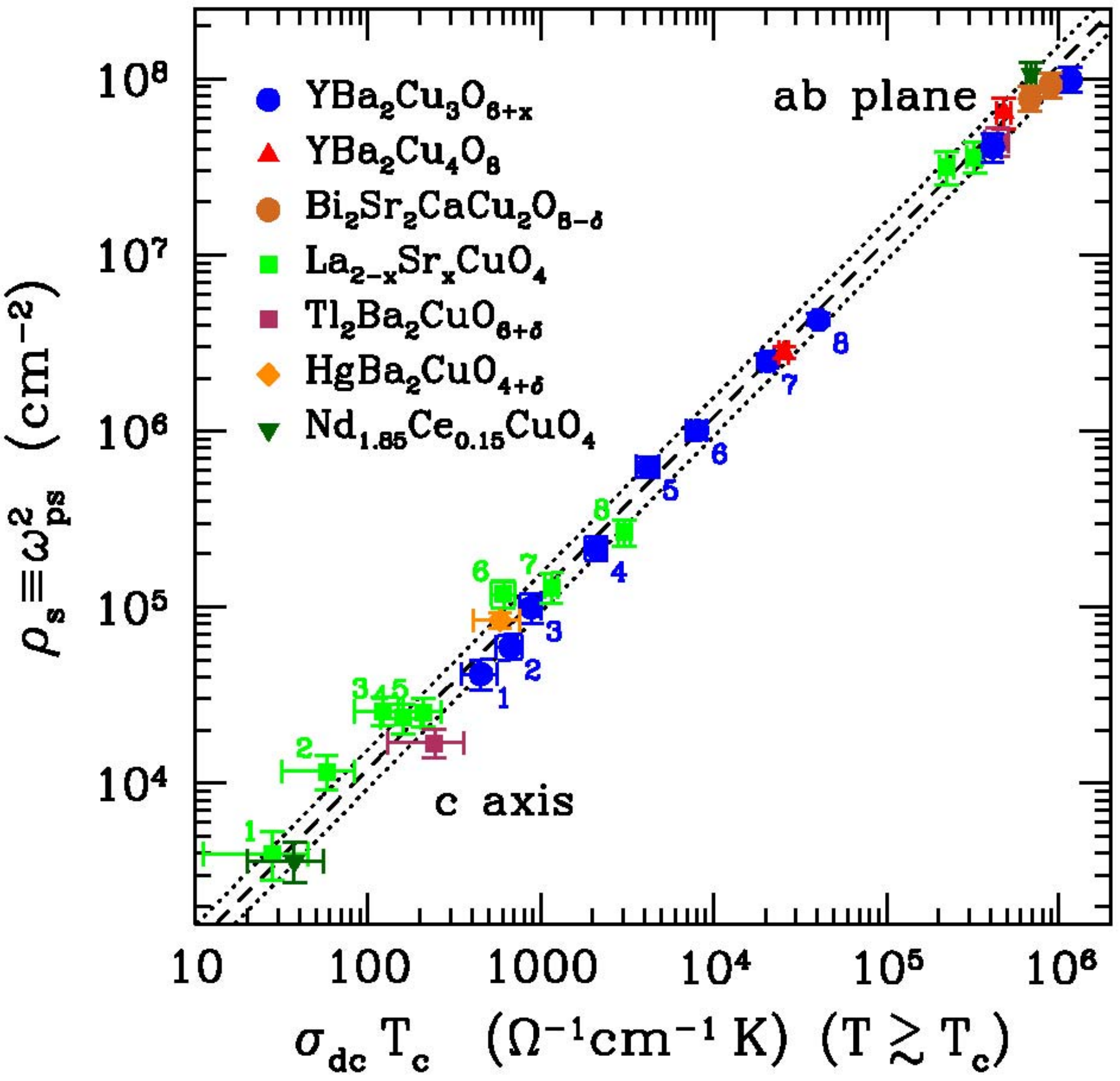

Figure 2: The log-log plot of the superfluid density $\rho_s \equiv \omega_{ps}^2$ vs $\sigma_{dc}$ $T_c$ for both the *a*-*b* planes and the *c* axis for a variety of cuprates (Tables 1 and 2). Within error, all of the data fall on the same universal (dashed) line with slope of unity, defined by $\rho_s = 120\ \sigma_{dc}\ T_c$; the dotted lines are from $\rho_s = (120 \pm 25)\ \sigma_{dc} T_c$.

Table 1: The critical temperature $T_c$, and the values parallel to the copper-oxygen (*a-b*) planes for the dc conductivity and the superfluid density for a variety of cuprate-based high-$T_c$ superconductors. The dc conductivity used in this scaling relation has been extrapolated from the optical conductivity $\sigma_{dc} = \sigma_1(\omega \rightarrow 0)$ at $T \approx T_c$. For $T << T_c$, the response of the dielectric function to the formation of a condensate is expressed purely by the real part $\varepsilon_1(\omega) = \varepsilon_\infty - \omega_{ps}^2/\omega^2$, which allows the strength of the condensate to be calculated from $\rho_s = -\omega^2 \varepsilon_1(\omega)$ in the $\omega \rightarrow 0$ limit. Here, $\omega_{ps}^2 = 4\pi n_s e^2/m^*$ is the square of the superconducting plasma frequency and $\rho_s \equiv \omega_{ps}^2$; the more familiar penetration depth $\lambda = 1/(2\pi\omega_{ps})$ is also shown. The values for the cuprates are compared with those of the simple metals Nb and Pb.

| Material | Ref. | $T_c$ (K) | $\sigma_{dc}$ ($\Omega^{-1}cm^{-1}$) | $\omega_{ps}$ ($cm^{-1}$) | $\lambda$ (μm) |
|---|---|---|---|---|---|
| $YBa_2Cu_3O_{6.60}$ | 10,11 | 59 | 6500 | 6400 | 0.24 |
| $YBa_2Cu_3O_{6.95}$ | 10,11 | 93.2 | 10500 | 9950 | 0.15 |
| Pr-$YBa_2Cu_3O_{7-\delta}$ | 12 | 40 | 2500 | 3700 | 0.43 |
| Pr-$YBa_2Cu_3O_{7-\delta}$ | 12 | 75 | 4900 | 7350 | 0.21 |
| $YBa_2Cu_3O_{7-\delta}$ | 12 | 92 | 8700 | 9200 | 0.17 |
| $YBa_2Cu_4O_8$ | 10 | 80 | 6000 | 8000 | 0.20 |
| $Bi_2Ca_2SrCu_2O_{8+\delta}$ | 12 | 85 | 8000 | 8800 | 0.18 |
| $Bi_2Ca_2SrCu_2O_{8+\delta}$ | 12 | 91 | 9800 | 9600 | 0.16 |
| Y/Pb-$Bi_2Ca_2SrCu_2O_{8+\delta}$ | 12 | 35 | 2500 | 3200 | 0.41 |
| Y-$Bi_2Ca_2SrCu_2O_{8+\delta}$ | 12 | 40 | 2600 | 3800 | 0.49 |
| $Tl_2Ba_2CuO_{6+\delta}$ | 13 | 88 | 5000 | 6630 | 0.24 |
| $Nd_{1.85}Ce_{0.15}CuO_4$ | 5,6 | 23 | 28000 | 10300 | 0.15 |
| $La_{1.87}Sr_{0.13}CuO_4$ | 14 | 33 | 7000 | 5600 | 0.28 |
| $La_{1.86}Sr_{0.14}CuO_4$ | 14 | 37 | 9000 | 6000 | 0.26 |
| Nb | 15 | 8.3 | 2.5e5 | 17600 | 0.09 |
| Nb | 16 | 9.3 | 8.5e5 | 35800 | 0.044 |
| Pb | 16 | 7.2 | 1.4e6 | 41000 | 0.038 |

Table 2: The critical temperature $T_c$, and the values perpendicular to the copper-oxygen planes (*c* axis) for the dc conductivity and the superfluid density in a variety of cuprate-based high-$T_c$ superconductors.

| Material | Ref. | Label | $T_c$ (K) | $\sigma_{dc}$ ($\Omega^{-1}cm^{-1}$) | $\omega_{ps}$ ($cm^{-1}$) | $\lambda$ ($\mu m$) |
|---|---|---|---|---|---|---|
| $YBa_2Cu_3O_{6.50}$ | 20 | 1 | 53 | 9 | 204 | 7.8 |
| $YBa_2Cu_3O_{6.60}$ | 20 | 2 | 58 | 12 | 244 | 6.5 |
| $YBa_2Cu_3O_{6.70}$ | 20 | 3 | 63 | 14 | 308 | 5.2 |
| $YBa_2Cu_3O_{6.80}$ | 20 | 4 | 78 | 27 | 465 | 3.4 |
| $YBa_2Cu_3O_{6.85}$ | 20 | 5 | 89 | 47 | 790 | 2.0 |
| $YBa_2Cu3O_{6.90}$ | 20 | 6 | 91.5 | 88 | 1003 | 1.6 |
| $YBa_2Cu3O_{6.95}$ | 20 | 7 | 93.2 | 220 | 1580 | 1.0 |
| $YBa2Cu3O_{6.99}$ | 21 | 8 | 90 | 450 | 2070 | 0.77 |
| | | | | | | |
| $YBa_2Cu_4O_8$ | 19 | | 80 | 320 | 1670 | 0.95 |
| $Tl_2Ba_2CuO_{6+\delta}$ | 22 | | 81 | 3 | 130 | 12.1 |
| $HgBa_2CuO_{4+\delta}$ | | | 97 | 6 | 290 | 5.5 |
| $Nd_{1.85}Ce_{0.15}CuO_4$ | 6 | | 23 | 1.5 | 60 | 27 |
| | | | | | | |
| $La_{1.92}Sr_{0.08}CuO_4$ | | 1 | 28 | 1.0 | 63 | 25 |
| $La_{1.90}Sr_{0.10}CuO_4$ | | 2 | 32 | 1.8 | 108 | 15 |
| $La_{1.88}Sr_{0.12}CuO_4$ | | 3 | 32 | 3.8 | 160 | 10 |
| $La_{1.875}Sr_{0.125}CuO_4$ | | 4 | 32 | 5.0 | 153 | 10 |
| $La_{1.875}Sr_{0.125}CuO_4$ | | 5 | 32 | 6.5 | 159 | 10 |
| $La_{1.85}Sr_{0.15}CuO_4$ | | 6 | 38 | 16 | 344 | 4.6 |
| $La_{1.83}Sr_{0.17}CuO_4$ | | 7 | 36 | 32 | 360 | 4.4 |
| $La_{1.80}Sr_{0.20}CuO_4$ | | 8 | 32 | 95 | 515 | 3.1 |